\documentclass[lettersize,journal]{IEEEtran}
\usepackage{amsmath,amsfonts}
\usepackage{algorithm}

\usepackage{algpseudocode}
\usepackage{xurl}

\usepackage{array}
\usepackage[caption=false,font=normalsize,labelfont=sf,textfont=sf]{subfig}
\usepackage{textcomp}

\usepackage{stfloats}
\usepackage{url}
\usepackage{verbatim}

\usepackage{graphicx}
\usepackage{multirow}
\usepackage{subcaption}  
\usepackage{subcaption}  
\usepackage{cite}
\usepackage{soul} 
\usepackage{placeins}

\usepackage{color} 
\usepackage{soul} 
\usepackage{booktabs}
\usepackage{tabularx}
\usepackage{orcidlink}

\usepackage{enumitem}          

\usepackage{tikz}
\usetikzlibrary{shapes, arrows, positioning}

\tikzstyle{block} = [rectangle, draw, fill=blue!20, 
    text width=6em, text centered, rounded corners, minimum height=2em, font=\small]
\tikzstyle{line} = [draw, -latex', thick]

\usepackage{hyperref}

\hypersetup{
    colorlinks=false,
    linkbordercolor=0 0 0,  
    citebordercolor=0 0 0   
}

\begin{document}

\title{Leveraging VAE-Derived Latent Spaces for Enhanced Malware Detection with Machine Learning Classifiers}

\author{Bamidele~Ajayi\orcidlink{0000-0003-1419-9375}\textsuperscript{1},
Basel~Barakat\orcidlink{0000-0001-9126-7613}\textsuperscript{2},
Ken~McGarry\orcidlink{0000-0002-9329-9835}\textsuperscript{1}\\[0.6ex]
\textsuperscript{1}School of Computer Science, University of Sunderland, Sunderland, UK\\
\textsuperscript{2}Department of Computing, Goldsmiths, University of London, UK}


\maketitle

\begin{abstract}

 This paper assesses the performance of five machine learning classifiers: Decision Tree, Naive Bayes, LightGBM, Logistic Regression, and Random Forest using latent representations learned by a Variational Autoencoder from malware datasets. Results from the experiments conducted on different training-test splits with different random seeds reveal that all the models perform well in detecting malware with ensemble methods (LightGBM and Random Forest) performing slightly better than the rest. In addition, the use of latent features reduces the computational cost of the model and the need for extensive hyperparameter tuning for improved efficiency of the model for deployment. Statistical tests show that these improvements are significant, and thus, the practical relevance of integrating latent space representation with traditional classifiers for effective malware detection in cybersecurity is established.

\end{abstract}

\section{Introduction} 

In today's hyperconnected world, malware attacks have risen to concerning proportions, presenting substantial challenges for cybersecurity. Sophisticated malware variants, such as viruses, worms, and ransomware, are progressively adept at circumventing traditional detection methods. The increasing complexity of these threats—spanning financial losses to critical infrastructure breaches—demands the creation of more resilient and adaptive strategies for malware detection and classification.

Conventional machine learning methods, like Decision Trees, Random Forests, and Support Vector Machines, have been extensively utilized in malware classification endeavours. These solutions generally depend on manually constructed features or statistical patterns extracted from the malicious code or behavioural data of malware samples \cite{souri2018}. Nonetheless, their efficacy is constrained by their incapacity to generalize to novel malware variants, especially in adversarial contexts. This constraint is exacerbated by malware developers employing obfuscation and polymorphism techniques, which modify the program's appearance while preserving its fundamental operation \cite{sihag2021}. These tactics undermine conventional methods, necessitating needs capable of identifying both known and novel malware.

Latent-space models, especially those constructed with deep learning frameworks such as Variational Autoencoders (VAEs), have presented itself as a viable solution to these issues. Variational Autoencoders (VAEs) are generative models that map input data into a reduced-dimensional latent space, preserving essential characteristics that may be obscured in the original high-dimensional space \cite{kingma2013}. This latent space helps dimensionality reduction and the identification of fundamental patterns crucial for efficient malware classification \cite{taylor2021}. The acquired latent representations can function as valuable characteristics for conventional machine learning algorithms, improving their ability to classify malware accurately despite obfuscation\cite{kim2022}.

This research presents a novel hybrid methodology that integrates Variational Autoencoders (VAEs) with traditional machine learning approaches, including Random Forests, Logistic Regression, Decision Trees, Naïve Bayes, and LightGBM across various configurations. We intend to overcome the shortcomings of conventional machine learning models and purely deep learning approaches by utilizing the latent space representations acquired via the VAE. This methodology integrates the advantages of deep learning in feature extraction and efficiency of traditional machine learning algorithms, providing a more adaptable and effective solution for malware detection.

A primary obstacle in utilizing latent space models for malware classification is the inherent imbalance found in the majority of malware datasets. Malware families are typically characterized by a few predominant kinds, whereas several infrequent variants constitute a minor segment of the dataset. This disparity affects the learning process, as models may find it challenging to appropriately represent infrequent virus variants \cite{zahoora2022}. Moreover, the prevalent application of obfuscation and polymorphism complicates the ability of conventional models to identify alterations in malware presentation without jeopardizing detection precision \cite{microage2024}. Our proposed VAE-based approach tackles these difficulties by transforming malware samples into a latent space for the extraction of invariant features, hence enhancing the efficacy of downstream classifiers in adversarial and dynamic settings.

A further advantage of utilizing latent space features is their interpretability. Conventional machine learning models frequently exhibit opaque decision-making processes, complicating practitioners' ability to understand the rationale behind certain decisions. Conversely, latent space representations provide a more rational understanding of the fundamental elements influencing classification decisions, facilitating more transparent and reliable systems for malware detection. Transparency is essential in critical cybersecurity applications, where knowing the reasoning behind a classification decision is as significant as the decision itself \cite{masud2024}.

This research presents a novel classification framework that combines latent space representations derived from a Variational Autoencoder (VAE) with traditional machine learning classifiers and various configurations. We pretrain the Variational Autoencoder (VAE) on a benchmark malware dataset\cite{EMBER} \cite{BODMAS}, then utilizing the acquired latent features as input for classifiers like Random Forests, Logistic Regression, Decision Trees, Naïve Bayes, and LightGBM. We conduct comprehensive tests to compare our method with leading classical and deep learning models, illustrating that the incorporation of latent space characteristics markedly enhances classification performance without requiring hyperparameter optimization.

This research key contributions are: We introduce an innovative hybrid method that utilizes VAE-derived latent space characteristics within traditional machine learning models for malware classification using various configurations. Secondly, we present a thorough empirical assessment of the influence of latent representations on classification with regards to processing efficiency, providing insights beneficial to both academia and practitioners in cybersecurity. 

By integrating latent space features with conventional machine learning algorithms, we offer a more efficient and comprehensible approach to malware identification. Our methodology not only improves performance and accuracy but also provides a viable strategy for addressing the continuously changing threat environment presented by sophisticated malware. This research explores new opportunities for utilizing latent space models in cybersecurity, enabling analysts to enhance the detection and classification of malware in the changing threat landscape.

\section{Related Work}

Investigating the possibilities for synthetic malware generation as opcode sequences using generative adversarial networks (GANs) and deep variational autoencoders (VAE). \cite {choi2024} focused on assessing synthetic malware in misleading machine learning classifiers, their research found that neither VAE nor GAN could effectively produce malware that avoids detection. Still, the WGAN-GP algorithm indicated promise in enhancing synthetic malware production since it needed more synthetic samples to get effective detection. Although the difficulty of avoiding complex classifiers remains major, this study emphasizes the increasing interest in generative models for malware generation.

For unsupervised malware identification \cite {kiran2024} presented the Hybrid Adversarial-variational Autoencoder (HAVAE), a model combining the strengths of Adversarial Autoencoders (AAEs) and VAEs. Without large-scale labelled datasets, HAVAE enhances the identification of malicious software by catching subtle elements inside the latent space. Showcasing strong performance across several datasets, the model generates realistic yet discriminative samples via the reparameterization method. Kiran's study highlights how well unsupervised learning detects malware and presents a creative way to feature extraction and classification in a changing threat environment.

With a Conditional Variational Autoencoder (CVAE) for Android malware family analysis,\cite {ban2022} overcame the constraints of imbalanced datasets in malware classification. Their method maintains malware's behaviour throughout data augmentation, therefore enhancing classification accuracy. A macro-F1 score of 0.91 and an accuracy of 0.99\% were obtained by the study showing that adding original malware set greatly improves the performance of the classifier. This work shows how generative models used with conventional datasets could overcome dataset imbalance and enhance malware detection.

Using Graph Variational Autoencoder (GVAE) to parse API-call graphs taken from Android APK files, \cite {gunduz2022} presented a new malware detection system. The work sought to lower graph node feature size and assess whether GVAE-reduced embeddings might improve malware detection performance. Showing notable gains, the GVAE was coupled with linear-based (SVM) and ensemble-based (LightGBM) systems. According to the study, GVAE-reduced embeddings roughly raised accuracy and F-measure rates for both models by 4\%. Furthermore, LightGBM with a smaller set of 30 features showed even more accuracy rate of 0.967 by combining recursive feature elimination (RFE) with GVAE embeddings, thereby stressing the efficiency of this hybrid approach. The contributions of the paper consist in the introduction of GVAE for feature reduction in malware detection, the application of sophisticated feature selection strategies, and validation of the suggested approaches by empirical results .

Transforming malware executables into image-based representations, \cite {kumar2021} suggested an Autoencoder Enhanced Deep Convolutional Neural Network (AE-DCNN) for malware classification. This method enhances classification performance by using a deep convolutional neural network (DCNN) with an encoder. On the Malimg dataset, the authors show a 10-fold cross-valuation accuracy of 99.38\% and an F1-score of 99.38\% proving the great efficacy of the technique. Eliminating conventional methods such feature engineering and reverse engineering, the AE-DCNN framework provides a more simplified and effective classification system. Furthermore, because of its texture-based study of malware files, the method demonstrates resistance to several obfuscation methods. By combining autoencoder improvements with deep learning methods, the AE-DCNN strategy marks a major breakthrough in malware classification and generates exceptional accuracy and resilience against obfuscation .

With an eye toward deriving a deep latent space that captures both feature and label interdependencies, \cite {yeh2017} present a new framework termed the Canonical-Correlated AutoEncoder (C2AE) for multi-label classification. To develop joint feature-label embeddings, their approach integrates Deep Canonical Correlation Analysis (DCCA) with an autoencoder architecture, hence improving the classification accuracy. The C2AE model can properly use label dependencies and manage missing labels during training by using a label-correlation aware loss function. In terms of accuracy and computational efficiency, tests spanning several datasets show that C2AE is superior over state-of- the-art multi-label classification techniques.

Because this work uses deep learning methods to embed labels into a latent space where associations between them are maintained, it is especially pertinent to latent space learning. Such methods have significant consequences for malware classification since different properties or traits could have overlapping or dependent behaviours. This approach is a valuable reference for applications needing effective classification with complicated dependencies since it underlines how latent spaces may be essential in lowering dimensionality while keeping important information for prediction tasks.

\cite {liu2018}, working on `Data Augmentation via Latent Space Interpolation for Image Classification,' The authors proposed to apply a uniform distribution on the latent space's feature representations by means of an Adversarial Autoencoder (AAE). Their more varied set of augmented data produced by using linear interpolation in this uniformly distributed latent space greatly enhanced the classification performance on benchmark datasets including ILSVRC 2012 and CIFAR-10.

This approach's main value is in its capacity to create reasonable and instructive training samples by interpolating across several latent representations, hence solving the "hole" issue sometimes present in high-dimensional latent spaces. This method not only extends the conventional data augmentation process but also offers a possible foundation for additional classification problems where producing reasonable alternative data points is essential.

By improving the usability of VAE-derived latent spaces as inputs to traditional machine learning models, we specifically addressed several important constraints noted in past methods. Regarding processing efficiency, accuracy, generalization, and the necessity of hyperparameter adjustment, this approach yields notable gains. The computational load related with processing high-dimensional data is much lowered by using the small latent space as input. For real-time or large-scale datasets, this reduces the training and inference phases of machine learning models, hence increasing their practicality.

\section{Mathematical Representation}\label{sec:vae_math}

\begin{enumerate}[label=\arabic*., leftmargin=*, itemsep=0pt]  
  \item \textbf{Encoder} – maps input $x$ to a latent Gaussian
  \[
    q_\phi(z\!\mid\!x)=\mathcal{N}\!\bigl(z;\,\mu_\phi(x),\operatorname{diag}\sigma_\phi^2(x)\bigr).
  \]

  \item \textbf{Decoder} – reconstructs $x$ from a latent sample
  \[
    p_\theta(x\!\mid\!z)=\mathcal{N}\!\bigl(x;\,\mu_\theta(z),\operatorname{diag}\sigma_\theta^2(z)\bigr).
  \]

  \item \textbf{Reparameterisation trick}
  \[
    z=\mu_\phi(x)+\sigma_\phi(x)\odot\epsilon,\qquad 
    \epsilon\sim\mathcal{N}(0,I).
  \]

  \item \textbf{Objective (ELBO)}
  \[
    \mathcal{L}(x;\theta,\phi)=
      \mathbb{E}_{q_\phi(z\mid x)}[\log p_\theta(x\mid z)]
      -D_{\mathrm{KL}}\!\bigl(q_\phi(z\mid x)\;\|\;p(z)\bigr).
  \]
\end{enumerate}

\begin{figure}[htbp]
\centering
\includegraphics[width=0.9\linewidth]{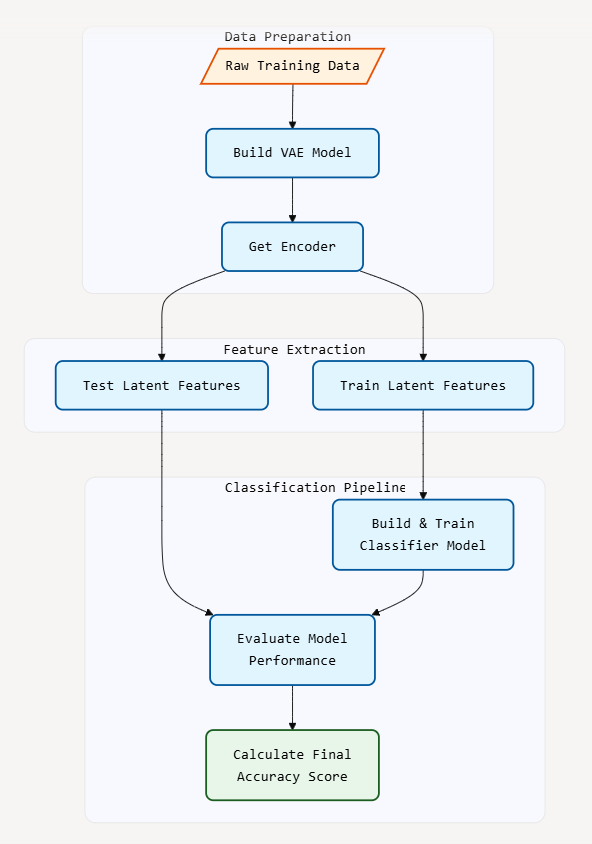}
\caption{Flow Diagram for Classifier with Latent Space Representations}
\label{fig:Flow Diagram for Classifier with Latent Space Representations}
\end{figure}

 We introduce a new algorithm that uses VAE latent space representations as features for traditional machine learning  classifiers. Unlike the classifiers trained on the raw input features, the models that use these learned latent representations with significant reduced dimension are comparably accurate and more robust. The VAE encodes the high-dimensional data into a low-dimensional latent  space, which is a compact and informative representation of the data that helps uncover the structure of the data  and improve the performance of the downstream classifier as shown in Figure~\ref{fig:Flow Diagram for Classifier with Latent Space Representations}.

\begin{algorithm}[H]
\caption{Classifier Evaluation with Original Features}
\label{alg:alg2}
\begin{algorithmic}[1]

\Function{Build}{data}
    \State $C \gets \text{YourClassifier()}$
    \State $C.\text{fit}(data.features, data.labels)$
    \State \textbf{return} $C$
\EndFunction

\Function{Evaluate}{C, test\_data}
    \State $pred \gets C.\text{predict}(test\_data.features)$
    \State $acc \gets \text{calc\_accuracy}(pred, test\_data.labels)$
    \State \textbf{return} $acc$
\EndFunction

\State $data \gets \text{LoadData()}$

\State $C \gets \text{Build}(data)$

\State $init\_acc \gets \text{Evaluate}(C, data.test\_data)$

\State \textbf{Print} ``Accuracy with all features: ", $init\_acc$

\end{algorithmic}
\end{algorithm}

\begin{algorithm}[H]
\caption{Classifier with Latent Space Representations}
\label{alg:alg1}
\begin{algorithmic}[1]

\Function{BuildVAE}{train\_data}
    \State $vae \gets \text{VAEModel()}$
    \State $vae.\text{compile}(optimizer=adam, loss=\text{vae\_loss})$
    \State $vae.\text{fit}(train\_data.features, train\_data.features, \newline epochs=50, batch\_size=64)$
    \State \textbf{return} $vae$
\EndFunction

\Function{ExtractLatent}{vae, data}
    \State $encoder \gets vae.\text{get\_encoder()}$
    \State $z\_mean \gets encoder.\text{predict}(data.features)$
    \State \textbf{return} $z\_mean$
\EndFunction

\Function{BuildClassifier}{latent\_data}
    \State $C \gets \text{YourClassifier()}$
    \State $C.\text{fit}(latent\_data, data.labels)$
    \State \textbf{return} $C$
\EndFunction

\Function{Evaluate}{C, test\_latent\_data}
    \State $pred \gets C.\text{predict}(test\_latent\_data)$
    \State $acc \gets \text{calc\_accuracy}(pred, test\_data.labels)$
    \State \textbf{return} $acc$
\EndFunction

\State $data \gets \text{LoadData()}$

\State $vae \gets \text{BuildVAE}(data.train)$

\State $train\_latent \gets \text{ExtractLatent}(vae, data.train)$
\State $test\_latent \gets \text{ExtractLatent}(vae, data.test)$

\State $C \gets \text{BuildClassifier}(train\_latent)$

\State $latent\_acc \gets \text{Evaluate}(C, test\_latent)$

\State \textbf{Print} ``Accuracy with latent features: ", $latent\_acc$

\end{algorithmic}
\end{algorithm}

\section{Methodology}

In order to assess the effectiveness of the latent space representations in the context of malware classification, two  phase experimental design was carried out in this paper. In the first phase, two benchmark malware datasets \cite{EMBER, BODMAS} were first normalized and then cleaned. The datasets which were initially dimensional feature vectors of size 2,381 were cleaned to exclude instances without labels and were divided into training, validation and test sets. More training-test splits were also made (30\%/30\%, 50\%/30\%, 70\%/30\%) to ensure that the model performance was evaluated with different levels of  data availability. All features were normalized between [0,1] range using MinMaxScaler to improve  convergence and enhance classifier performance since features were on different scales.

In the next phase, a Variational  Autoencoder (VAE) was trained on the preprocessed training data to learn compact latent features.  The encoder was the dense layers that reduced the dimensionality to a 32 dimensional latent space of mean (z\_mean) and log variance (z\_log\_var). The latent vectors were generated using a reparameterization trick to ensure that  the latent space was learned smoothly while the decoder attempted to reconstruct the original input. The VAE was trained for 50 iterations with a batch size of 64,  using a custom loss function that incorporated Mean Squared Error for reconstruction loss and  Kullback-Leibler divergence for regularization.

The latent representations learned by the VAE were then employed as input features to  five different machine learning classifiers, including Decision Tree, Naive Bayes, LightGBM, Logistic  Regression, and Random Forest. These models were chosen as representative of different approaches to learning and known to  perform well in malware detection tasks. Each classifier was trained with five-fold cross validation, two different random  seeds (42 and 123) and accuracy and Area Under the Curve (AUC) were used  as performance metrics to evaluate the models on the test set. T-tests were used to determine the statistical  significance of the performance differences to confirm that the observed performance differences were robust. Furthermore, confusion matrices,  classification reports and ROC curves were constructed to give a more complete view of the performance, and execution times  were recorded to judge the computational efficiency.

With the help of the unsupervised feature learning component based  on the VAE and the traditional classifiers, the detection accuracy is improved while the computational cost and the  need for extensive hyperparameter tuning are reduced. All the models that were trained in this research, including  the VAE encoder and the classifiers, have been saved in order to ensure reproducibility. The  complete Python code for data preprocessing, VAE training, the extraction of the latent space, and the  training of the classifiers is provided in the supplementary material to enable further research into machine learning based malware  detection.

\section{Evaluation and Results}

 The experiments rigorously evaluate model accuracy across different training-test split ratios (30/30, 50/30, and 70/30) and random seeds (42 and 123) to assess robustness. A key novelty of this approach is that no hyperparameter tuning was required, yet the models achieved high performance directly from latent space features, demonstrating the inherent advantages of this representation when compared to \ref{tab:model_performance_no_hyperparameter_tuning_ember} 
\ref{tab:model_performance_hyperparameter_tuning_ember}
 \ref{tab:model_performance_with_no_hyperparameter_tuning_bodmas}
 \ref{tab:model_performance_with_hyperparameter_tuning_bodmas} .
Decision Tree classifiers had mean cross-validation scores of ~0.931 and test accuracies around 95\%. Although results varied with split ratios, they remained stable across different random seeds. The robustness of Decision Trees across varying conditions suggests that their hierarchical structure allows them to generalize well within the latent space. Importantly, latent space features improved memory consumption and inference speed, making this method highly efficient compared to training on high-dimensional raw data or full feature sets as shown in Table \ref {tab:bodmas-classifier-performance-bodmas}.\par
Naive Bayes, while not as accurate as tree-based models, demonstrated efficiency in ranking malware classes. In BODMAS, its cross-validation scores decreased from 0.7967 for the 30/30 split to 0.759 for the 70/30 split, with test accuracy following the same trend. EMBER showed similar results as shown in Table \ref {tab:bodmas-classifier-performance-bodmas}. Despite lower accuracy, high AUC values confirmed its usefulness for ranking, and its rapid inference time and low computational complexity make it particularly valuable in resource-constrained environments. The observed drop in performance with increased training size suggests that Naive Bayes might struggle with increased complexity in feature representations derived from latent space.\par
LightGBM demonstrated the highest accuracy, achieving cross-validation scores of 0.9388 and test accuracy of 0.9399 in BODMAS. Its AUC values exceeded 0.947 across all splits, confirming its effectiveness . The application of LightGBM in latent space significantly improved inference speed and reduced memory consumption, proving its suitability for real-world deployment. However, LightGBM exhibited variability across random states, indicating that its sensitivity to initialization might require further tuning or ensemble techniques to enhance stability.\par
\begin{figure}[htbp]
\centering
\includegraphics[width=0.8\linewidth]{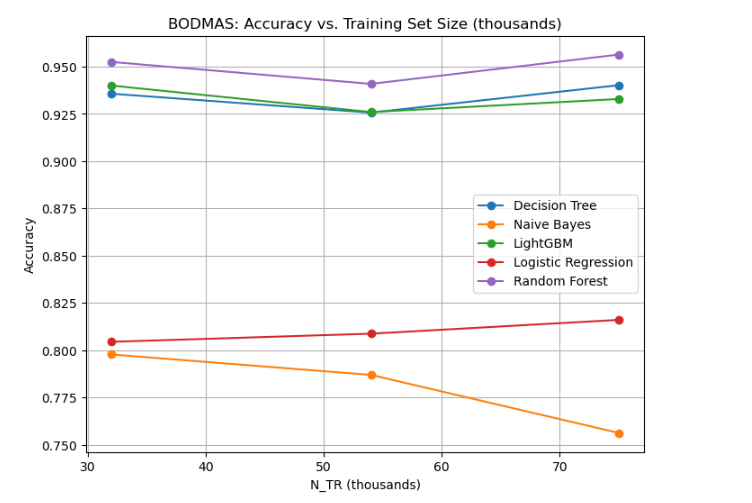}
\caption{BODMAS Latent Feature Accuracy VS Training}
\label{fig:BODMAS Latent Feature Accuracy VS Training}
\end{figure}
\begin{figure}[htbp]
\centering
\includegraphics[width=0.8\linewidth]{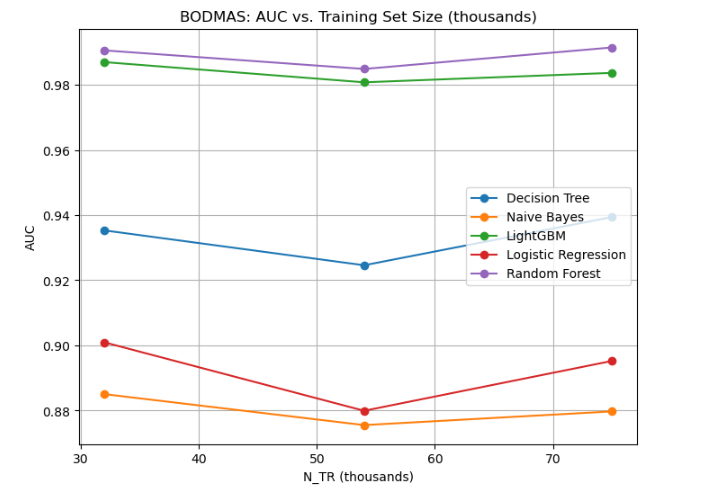}
\caption{BODMAS Latent Feature AUC VS Training.}
\label{fig:BODMAS Latent Feature AUC VS Training}
\end{figure}

Logistic Regression, though stable, was outperformed by ensemble methods. It had cross-validation scores of ~0.7996, test accuracies near 83\%, and an AUC of 0.9009. While limited by linear decision boundaries, its efficiency in latent space representation makes it an excellent choice for low-latency applications. The lack of significant performance variation across splits further underscores its reliability, particularly in real-time threat detection where computational overhead is a concern as shown in \ref{fig:BODMAS Latent Feature Accuracy VS Training}
 \ref{fig:BODMAS Latent Feature AUC VS Training} \ref{tab:bodmas-classifier-performance-bodmas}.\par
\begin{figure}[htbp]
\centering
\includegraphics[width=0.8\linewidth]{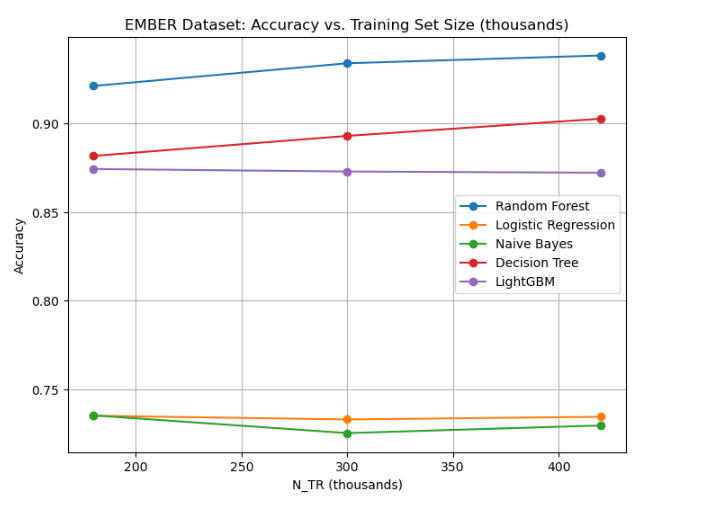}
\caption{EMBER Latent Feature Accuracy VS Training}
\label{fig:EMBER Latent Feature Accuracy VS Training}
\end{figure}
\begin{figure}[htbp]
\centering
\includegraphics[width=0.8\linewidth]{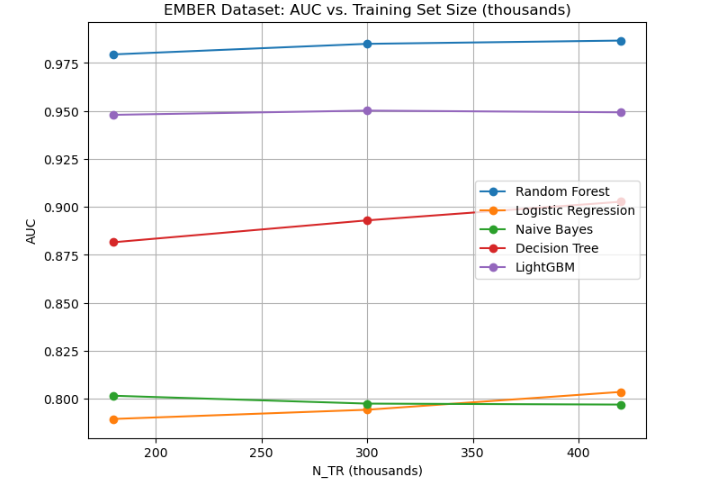}
\caption{EMBER Latent Feature AUC VS Training}
\label{fig:EMBER Latent Feature AUC VS Training}
\end{figure}

Random Forest exhibited the best classification performance overall. With a 30/30 split, it achieved cross-validation scores of 0.9498, test accuracy of 0.9523, and an AUC close to 0.9906. Similar results were observed for EMBER. The ensemble nature of Random Forest allows it to capture complex relationships within latent space features, leading to high accuracy. However, statistically significant differences between the 30/30 and other splits suggest that Random Forest benefits from larger training datasets, reinforcing the importance of training-test data proportions as shown in \ref{tab:ember-classifier-performance-ember} \ref{fig:EMBER Latent Feature AUC VS Training} \ref{fig:EMBER Latent Feature Accuracy VS Training}.\par
The comparison of different random states (42 vs. 123) reveals that for most classifiers, the t-statistics are very close to zero and the p-values are extremely high (e.g., p = 0.8072, 1.0000), indicating no statistically significant difference in performance. This suggests that for both BODMAS and EMBER datasets, classifier outcomes are robust to the choice of random seed. However, an exception appears with the Random Forest classifier in the EMBER 50/30 split, where the t-statistic is -6.6940 and the p-value is 0.0002, indicating sensitivity to the random state in this particular instance. This could be due to an interaction between the model’s randomness and the specific data split.
When analyzing the effect of data splits, BODMAS comparisons across split ratios (30/30 vs. 50/30, 30/30 vs. 70/30, and 50/30 vs. 70/30) exhibit large t-statistics and consistently low p-values ($p < 0.05$), showing that changing the split ratios significantly alters classifier performance. This demonstrates that the proportion of data allocated to training versus testing has a significant impact on results as shown in \ref{tab:random_states_comparison}
\ref{tab:split_comparison}.\par In the EMBER dataset, similar significant differences are found for most classifiers, confirming that performance is affected by the split ratio. Logistic Regression remains an exception, showing no significant differences across splits (p-values between 0.26 and 0.94), suggesting stability regardless of how data is partitioned. LightGBM displays mixed sensitivity to split ratios, with significance depending on the specific comparison and random state.
T-tests further confirm that most classifiers had statistically similar cross-validation scores across random seeds ($p < 0.05$), except for Random Forest on EMBER 50/30 (t = -6.6940, p = 0.0002). Comparisons across splits highlight significant differences for Decision Tree, Naive Bayes, and Random Forest, while Logistic Regression remains stable, emphasizing the critical impact of split sensitivity on classifier performance. These results indicate that the method of data partitioning plays a critical role, as significant differences in performance are observed across different split ratios as shown in \ref{tab:ember_random_states} \ref{tab:ember_split_comparison}.
\begin{table}[H]
\centering
\resizebox{\columnwidth}{!}{
\begin{tabular}{llcc}
\toprule
\textbf{Split} & \textbf{Classifier} & \textbf{t-statistic} & \textbf{p-value} \\
\midrule
EMBER 30/30 & Decision Tree       & 0.0754  & 0.9418 \\
           & Naive Bayes         & 0.0000  & 1.0000 \\
           & LightGBM            & 0.0000  & 1.0000 \\
           & Logistic Regression & 0.0000  & 1.0000 \\
           & Random Forest       & -0.3691 & 0.7216 \\
\midrule
EMBER 50/30 & Decision Tree       & -0.0477 & 0.9631 \\
           & Naive Bayes         & 0.0000  & 1.0000 \\
           & LightGBM            & -0.6680 & 0.5229 \\
           & Logistic Regression & 0.0000  & 1.0000 \\
           & Random Forest       & -6.6940 & 0.0002 \\
\midrule
EMBER 70/30 & Decision Tree       & -0.4554 & 0.6609 \\
           & Naive Bayes         & 0.0000  & 1.0000 \\
           & LightGBM            & -0.2266 & 0.8264 \\
           & Logistic Regression & 0.0000  & 1.0000 \\
           & Random Forest       & 0.0168  & 0.9870 \\
\bottomrule
\end{tabular}
}
\caption{Comparison of random states (42 vs 123) for each classifier within each EMBER split.}
\label{tab:ember_random_states}
\end{table}
\begin{table}[H]
\centering
\resizebox{\columnwidth}{!}{
\begin{tabular}{llccc}
\toprule
\textbf{Classifier} & \textbf{Random State} & \textbf{Comparison} & \textbf{t-statistic} & \textbf{p-value} \\
\midrule
\multirow{3}{*}{Decision Tree} & \multirow{3}{*}{42} & 30/30 vs 50/30 & -16.1213 & 0.0000 \\
                             &                    & 30/30 vs 70/30 & -28.7708 & 0.0000 \\
                             &                    & 50/30 vs 70/30 & -12.2503 & 0.0000 \\
\midrule
\multirow{3}{*}{Decision Tree} & \multirow{3}{*}{123} & 30/30 vs 50/30 & -13.8541 & 0.0000 \\
                             &                     & 30/30 vs 70/30 & -25.6003 & 0.0000 \\
                             &                     & 50/30 vs 70/30 & -10.1159 & 0.0000 \\
\midrule
\multirow{3}{*}{Naive Bayes}   & \multirow{3}{*}{42}  & 30/30 vs 50/30 & 7.7939  & 0.0001 \\
                             &                     & 30/30 vs 70/30 & 4.7865  & 0.0014 \\
                             &                     & 50/30 vs 70/30 & -3.8106 & 0.0052 \\
\midrule
\multirow{3}{*}{Naive Bayes}   & \multirow{3}{*}{123} & 30/30 vs 50/30 & 7.7939  & 0.0001 \\
                             &                     & 30/30 vs 70/30 & 4.7865  & 0.0014 \\
                             &                     & 50/30 vs 70/30 & -3.8106 & 0.0052 \\
\midrule
\multirow{3}{*}{LightGBM}      & \multirow{3}{*}{42}  & 30/30 vs 50/30 & -2.2882 & 0.0514 \\
                             &                     & 30/30 vs 70/30 & -0.6418 & 0.5390 \\
                             &                     & 50/30 vs 70/30 & 1.3736  & 0.2068 \\
\midrule
\multirow{3}{*}{LightGBM}      & \multirow{3}{*}{123} & 30/30 vs 50/30 & -4.2594 & 0.0028 \\
                             &                     & 30/30 vs 70/30 & -1.1109 & 0.2989 \\
                             &                     & 50/30 vs 70/30 & 2.8681  & 0.0209 \\
\midrule
\multirow{3}{*}{Logistic Regression} & \multirow{3}{*}{42} & 30/30 vs 50/30 & 1.2077 & 0.2616 \\
                                    &                     & 30/30 vs 70/30 & 0.0717 & 0.9446 \\
                                    &                     & 50/30 vs 70/30 & -0.9205 & 0.3842 \\
\midrule
\multirow{3}{*}{Logistic Regression} & \multirow{3}{*}{123} & 30/30 vs 50/30 & 1.2077 & 0.2616 \\
                                    &                     & 30/30 vs 70/30 & 0.0717 & 0.9446 \\
                                    &                     & 50/30 vs 70/30 & -0.9205 & 0.3842 \\
\midrule
\multirow{3}{*}{Random Forest} & \multirow{3}{*}{42}  & 30/30 vs 50/30 & -25.5407 & 0.0000 \\
                             &                     & 30/30 vs 70/30 & -42.0416 & 0.0000 \\
                             &                     & 50/30 vs 70/30 & -6.7869  & 0.0001 \\
\midrule
\multirow{3}{*}{Random Forest} & \multirow{3}{*}{123} & 30/30 vs 50/30 & -40.6750 & 0.0000 \\
                             &                     & 30/30 vs 70/30 & -40.6750 & 0.0000 \\
                             &                     & 50/30 vs 70/30 & 0.0000   & 1.0000 \\
\bottomrule
\end{tabular}
}
\caption{Comparison of splits for each classifier for each random state within EMBER.}
\label{tab:ember_split_comparison}
\end{table}

Key insights from this evaluation include:
\begin{itemize}
\item \textbf  Most classifiers exhibited consistency across random seeds, reinforcing the reliability of the evaluation approach.
\item \textbf Data partitioning significantly affected the performance of Decision Tree and Naive Bayes, emphasizing the necessity of careful data allocation.
\item \textbf Random Forest and LightGBM outperformed other models, particularly benefiting from larger training datasets.
\item \textbf The ability to achieve high classification performance without hyperparameter tuning demonstrates a novel and efficient approach for malware classification.
\end{itemize}
These findings have far-reaching implications for large-scale cybersecurity applications. The proposed methodology enables malware detection with reduced inference time and lower memory consumption, making it particularly suitable for security frameworks, SIEM, and EDR tools. Additionally, the lightweight computational cost of latent space-based classification makes it highly viable for IoT security and edge computing. Security vendors can integrate these techniques into their real-time malware detection and response mechanisms to enhance efficiency and scalability as shown in \ref{tab:execution-times}.

This research introduces a novel methodology demonstrating that latent space representations enhance malware classification accuracy while reducing computational overhead. The integration of Random Forest and LightGBM into cybersecurity frameworks can significantly improve real-time threat detection in constrained environments. The high statistical significance of these findings underscores their generalizability, offering a groundbreaking, low-overhead approach to malware classification that can be seamlessly deployed across various cybersecurity applications.

\begin{table}[htbp]
\centering
\resizebox{\columnwidth}{!}{
\begin{tabular}{lllllll}
\toprule
\multicolumn{7}{c}{\textbf{Performance Metrics (EMBER)}} \\
\midrule
$\mathbf{N_{TR}}$ & $\mathbf{N_{TS}}$ & \textbf{Classifier} & \textbf{AUC} & \textbf{Accuracy} & \textbf{Mean CV Score} & \textbf{Std Dev} \\
($\mathbf{10^3}$) & ($\mathbf{10^3}$) &  &  &  &  &  \\
\midrule
\multicolumn{7}{c}{\textbf{30\% Training, 30\% Testing, 40\% Holdout}} \\
\midrule
180 & 180 & Random Forest        & 0.9794 & 0.9211 & 0.9156 & 0.0009 \\
180 & 180 & Logistic Regression  & 0.7894 & 0.7349 & 0.7346 & 0.0021 \\
180 & 180 & Naive Bayes          & 0.8015 & 0.7351 & 0.7355 & 0.0018 \\
180 & 180 & Decision Tree        & 0.8815 & 0.8816 & 0.8742 & 0.0016 \\
180 & 180 & LightGBM             & 0.9479 & 0.8742 & 0.8715 & 0.0011 \\
\midrule
\multicolumn{7}{c}{\textbf{50\% Training, 30\% Testing, 20\% Holdout}} \\
\midrule
300 & 180 & Random Forest        & 0.9849 & 0.9339 & 0.9321 & 0.0010 \\
300 & 180 & Logistic Regression  & 0.7942 & 0.7328 & 0.7331 & 0.0013 \\
300 & 180 & Naive Bayes          & 0.7974 & 0.7251 & 0.7263 & 0.0015 \\
300 & 180 & Decision Tree        & 0.8929 & 0.8929 & 0.8902 & 0.0012 \\
300 & 180 & LightGBM             & 0.9501 & 0.8728 & 0.8737 & 0.0016 \\
\midrule
\multicolumn{7}{c}{\textbf{70\% Training, 30\% Testing}} \\
\midrule
420 & 180 & Random Forest        & 0.9866 & 0.9383 & 0.9356 & 0.0004 \\
420 & 180 & Logistic Regression  & 0.8035 & 0.7343 & 0.7345 & 0.0026 \\
420 & 180 & Naive Bayes          & 0.7969 & 0.7294 & 0.7302 & 0.0013 \\
420 & 180 & Decision Tree        & 0.9026 & 0.9026 & 0.8988 & 0.0007 \\
420 & 180 & LightGBM             & 0.9492 & 0.8721 & 0.8724 & 0.0011 \\
\bottomrule
\end{tabular}
}
\caption{Performance of Various Classifiers on EMBER Dataset Splits}
\label{tab:ember-classifier-performance-ember}
\end{table}

\begin{table}[htbp]
\centering
\resizebox{\columnwidth}{!}{
\begin{tabular}{lllllll}
\toprule
\multicolumn{7}{c}{\textbf{Performance Metrics (BODMAS)}} \\
\midrule
$\mathbf{N_{TR}}$ & $\mathbf{N_{TS}}$ & \textbf{Classifier} & \textbf{AUC} & \textbf{Accuracy} & \textbf{Mean CV Score} & \textbf{Std Dev} \\
($\mathbf{10^3}$) & ($\mathbf{10^3}$) &  &  &  &  &  \\
\midrule
\multicolumn{7}{c}{\textbf{30\% Training, 30\% Testing, 40\% Holdout}} \\
\midrule
32 & 40 & Decision Tree        & 0.9353 & 0.9356 & 0.9314 & 0.0021 \\
32 & 40 & Naive Bayes          & 0.8850 & 0.7977 & 0.7967 & 0.0054 \\
32 & 40 & LightGBM             & 0.9870 & 0.9399 & 0.9388 & 0.0027 \\
32 & 40 & Logistic Regression  & 0.9009 & 0.8044 & 0.7996 & 0.0027 \\
32 & 40 & Random Forest        & 0.9906 & 0.9524 & 0.9498 & 0.0026 \\
\midrule
\multicolumn{7}{c}{\textbf{50\% Training, 30\% Testing, 20\% Holdout}} \\
\midrule
54 & 67 & Decision Tree        & 0.9246 & 0.9256 & 0.9238 & 0.0012 \\
54 & 67 & Naive Bayes          & 0.8755 & 0.7869 & 0.7892 & 0.0010 \\
54 & 67 & LightGBM             & 0.9808 & 0.9259 & 0.9274 & 0.0014 \\
54 & 67 & Logistic Regression  & 0.8799 & 0.8087 & 0.8063 & 0.0018 \\
54 & 67 & Random Forest        & 0.9849 & 0.9408 & 0.9404 & 0.0017 \\
\midrule
\multicolumn{7}{c}{\textbf{70\% Training, 30\% Testing}} \\
\midrule
75 & 94 & Decision Tree        & 0.9394 & 0.9401 & 0.9391 & 0.0012 \\
75 & 94 & Naive Bayes          & 0.8797 & 0.7562 & 0.7592 & 0.0042 \\
75 & 94 & LightGBM             & 0.9837 & 0.9328 & 0.9355 & 0.0005 \\
75 & 94 & Logistic Regression  & 0.8952 & 0.8160 & 0.8105 & 0.0023 \\
75 & 94 & Random Forest        & 0.9915 & 0.9563 & 0.9542 & 0.0005 \\
\bottomrule
\end{tabular}
}
\caption{Performance of Various Classifiers on the BODMAS Dataset Splits}
\label{tab:bodmas-classifier-performance-bodmas}
\end{table}

\begin{table}[htbp]
\centering
\resizebox{\columnwidth}{!}{
\begin{tabular}{lllllll}
\toprule
\textbf{Model} & \textbf{No Compoenents} & \textbf{Accuracy} & \textbf{Precision} & \textbf{Recall} & \textbf{F1 Score} & \textbf{ROC AUC} \\ \midrule
Random Forest &ALL & 0.9603            & 0.9674             & 0.9527           & 0.9600            & 0.9940 \\ \midrule
Logistic Regression &ALL & 0.5564            & 0.5414             & 0.7339           & 0.6231            & 0.6133 \\ \midrule
Decision Tree &ALL & 0.9394            & 0.9362             & 0.9430           & 0.9396            & 0.9394 \\ \midrule
Naive Bayes &ALL & 0.5346            & 0.9098             & 0.0761           & 0.1405            & 0.6075 \\ \midrule
Light GBM &ALL & 0.9516            & 0.9513             & 0.9518           & 0.9515            & 0.9907 \\  \bottomrule \\
\end{tabular}
}
\caption{EMBER Performance metrics for models with complete feature and no hyperparameter tuning}
\label{tab:model_performance_no_hyperparameter_tuning_ember}
\end{table}

\begin{table}[htbp]
\centering
\resizebox{\columnwidth}{!}{
\begin{tabular}{lllllll}
\toprule
\textbf{Model} & \textbf{No Components}  & \textbf{Accuracy} & \textbf{Precision} & \textbf{Recall} & \textbf{F1 Score} & \textbf{ROC AUC} \\ \midrule
Light GBM &ALL & 0.9807 & 0.9840 & 0.9773 & 0.9807 & 0.9979 \\ \midrule
Logistic Regression &ALL & 0.8872 & 0.8759 & 0.9021 & 0.8888 & 0.9589 \\ \midrule
Naive Bayes &ALL & 0.5346 & 0.9098 & 0.0761 & 0.1405 & 0.6075 \\ \midrule
Random Forest &ALL & 0.9628 & 0.9701 & 0.9550 & 0.9625 & 0.9947 \\ \midrule
Decision Tree &ALL & 0.9176 & 0.9206 & 0.9139 & 0.9172 & 0.9271 \\ 
\bottomrule \\
\end{tabular}
}
\caption{EMBER Performance metrics for models with complete feature and hyperparameter tuning}
\label{tab:model_performance_hyperparameter_tuning_ember}
\end{table}

\begin{table}[htbp]
\centering
\resizebox{\columnwidth}{!}{
\begin{tabular}{lllllll}
\toprule
\textbf{Model} & \textbf{No Componentss} & \textbf{Accuracy} & \textbf{Precision} & \textbf{Recall} & \textbf{F1 Score} & \textbf{ROC AUC} \\ \midrule
Logistic Regression &ALL & 0.7466 & 0.6642 & 0.8206 & 0.7342 & 0.8457 \\ \midrule
LightGBM &ALL & 0.9959 & 0.9953 & 0.9951 & 0.9952 & 0.9998 \\ \midrule
Decision Tree &ALL & 0.9888 & 0.9847 & 0.9892 & 0.9869 & 0.9889 \\ \midrule
Random Forest &ALL & 0.9945 & 0.9974 & 0.9896 & 0.9935 & 0.9997  \\ \midrule
Naive Bayes &ALL & 0.4920 & 0.4547 & 0.9605 & 0.6172 & 0.5614 \\ 
\bottomrule \\
\end{tabular}
}
\caption{BODMAS Performance metrics for models with complete feature and no hyperparameter tuning}
\label{tab:model_performance_with_no_hyperparameter_tuning_bodmas}
\end{table}

\begin{table}[H]
\centering
\resizebox{\columnwidth}{!}{
\begin{tabular}{lllllll}
\toprule
\textbf{Model} & \textbf{No Components} & \textbf{Accuracy} & \textbf{Precision} & \textbf{Recall} & \textbf{F1 Score} &\textbf{ROC AUC} \\ \midrule
Decision Tree &ALL  & 0.9881 & 0.9833 & 0.9887 & 0.9860 & 0.9881 \\ \midrule
LightGBM & ALL & 0.9958 & 0.9947 & 0.9954 & 0.9950 & 0.9998 \\ \midrule
Logistic Regression &ALL & 0.7482 & 0.6682 & 0.8133 & 0.7337 & 0.8428 \\ \midrule
Naive Bayes&ALL & 0.4920 & 0.4547 & 0.9605 & 0.6172 & 0.5614 \\ \midrule
Random Forest &ALL & 0.9945 & 0.9975 & 0.9896 & 0.9935 & 0.9997 \\ \bottomrule \\
\end{tabular}
}
\caption{BODMAS Comparison of models with complete feature and hyperparameter tuning}
\label{tab:model_performance_with_hyperparameter_tuning_bodmas}
\end{table}

\begin{table}[H]
\centering
\resizebox{\columnwidth}{!}{
\begin{tabular}{llcc}
\toprule
\textbf{Split} & \textbf{Classifier} & \textbf{t-statistic} & \textbf{p-value} \\
\midrule
BODMAS 30/30 & Decision Tree         & -0.2523 & 0.8072 \\
            & Naive Bayes           & 0.0000  & 1.0000 \\
            & LightGBM              & 0.0000  & 1.0000 \\
            & Logistic Regression   & 0.0000  & 1.0000 \\
            & Random Forest         & 0.4842  & 0.6413 \\
\midrule
BODMAS 50/30 & Decision Tree         & 0.0188  & 0.9855 \\
            & Naive Bayes           & 0.0000  & 1.0000 \\
            & LightGBM              & 0.0000  & 1.0000 \\
            & Logistic Regression   & 0.0000  & 1.0000 \\
            & Random Forest         & 0.2128  & 0.8368 \\
\midrule
BODMAS 70/30 & Decision Tree         & 0.4778  & 0.6456 \\
            & Naive Bayes           & 0.0000  & 1.0000 \\
            & LightGBM              & 0.0000  & 1.0000 \\
            & Logistic Regression   & 0.0000  & 1.0000 \\
            & Random Forest         & -1.0562 & 0.3217 \\
\bottomrule
\end{tabular}
}
\caption{Comparison of random states (42 vs 123) for each classifier within each split.}
\label{tab:random_states_comparison}
\end{table}

\begin{table}[H]
\centering
\resizebox{\columnwidth}{!}{
\begin{tabular}{llccc}
\toprule
\textbf{Classifier} & \textbf{Random State} & \textbf{Comparison} & \textbf{t-statistic} & \textbf{p-value} \\
\midrule
\multirow{3}{*}{Decision Tree} & \multirow{3}{*}{42} & 30/30 vs 50/30 & 6.1411  & 0.0003 \\
                               &                     & 30/30 vs 70/30 & -6.2678 & 0.0002 \\
                               &                     & 50/30 vs 70/30 & -17.6574& 0.0000 \\
\midrule
\multirow{3}{*}{Decision Tree} & \multirow{3}{*}{123} & 30/30 vs 50/30 & 6.9828  & 0.0001 \\
                               &                      & 30/30 vs 70/30 & -6.4404 & 0.0002 \\
                               &                      & 50/30 vs 70/30 & -23.1574& 0.0000 \\
\midrule
\multirow{3}{*}{Naive Bayes}   & \multirow{3}{*}{42}  & 30/30 vs 50/30 & 2.7497  & 0.0251 \\
                               &                     & 30/30 vs 70/30 & 10.9515 & 0.0000 \\
                               &                     & 50/30 vs 70/30 & 13.7383 & 0.0000 \\
\midrule
\multirow{3}{*}{Naive Bayes}   & \multirow{3}{*}{123} & 30/30 vs 50/30 & 2.7497  & 0.0251 \\
                               &                      & 30/30 vs 70/30 & 10.9515 & 0.0000 \\
                               &                      & 50/30 vs 70/30 & 13.7383 & 0.0000 \\
\midrule
\multirow{3}{*}{LightGBM}      & \multirow{3}{*}{42}  & 30/30 vs 50/30 & 7.5691  & 0.0001 \\
                               &                     & 30/30 vs 70/30 & 2.4185  & 0.0419 \\
                               &                     & 50/30 vs 70/30 & -11.1428& 0.0000 \\
\midrule
\multirow{3}{*}{LightGBM}      & \multirow{3}{*}{123} & 30/30 vs 50/30 & 7.5691  & 0.0001 \\
                               &                      & 30/30 vs 70/30 & 2.4185  & 0.0419 \\
                               &                      & 50/30 vs 70/30 & -11.1428& 0.0000 \\
\midrule
\multirow{3}{*}{Logistic Regression} & \multirow{3}{*}{42} & 30/30 vs 50/30 & -4.1039 & 0.0034 \\
                                    &                      & 30/30 vs 70/30 & -6.0987 & 0.0003 \\
                                    &                      & 50/30 vs 70/30 & -2.8430 & 0.0217 \\
\midrule
\multirow{3}{*}{Logistic Regression} & \multirow{3}{*}{123} & 30/30 vs 50/30 & -4.1039 & 0.0034 \\
                                    &                      & 30/30 vs 70/30 & -6.0987 & 0.0003 \\
                                    &                      & 50/30 vs 70/30 & -2.8430 & 0.0217 \\
\midrule
\multirow{3}{*}{Random Forest}  & \multirow{3}{*}{42}  & 30/30 vs 50/30 & 5.9883  & 0.0003 \\
                               &                     & 30/30 vs 70/30 & -3.2663 & 0.0114 \\
                               &                     & 50/30 vs 70/30 & -15.2201& 0.0000 \\
\midrule
\multirow{3}{*}{Random Forest}  & \multirow{3}{*}{123} & 30/30 vs 50/30 & 5.5212  & 0.0006 \\
                               &                      & 30/30 vs 70/30 & -6.3946 & 0.0002 \\
                               &                      & 50/30 vs 70/30 & -9.9289 & 0.0000 \\
\bottomrule
\end{tabular}
}
\caption{Comparison of splits for each classifier for each random state within BODMAS.}
\label{tab:split_comparison}
\end{table}

\section{Conclusions}

 This research proves that by using latent space representations, it is possible to improve the malware classification accuracy  without increasing the computational complexity of the model. Out of all the classifiers discussed, ensemble methods such as  Random Forest and LightGBM performed best, attaining highest accuracy values and AUC scores in all  the dataset splits and random seeds. These results are also statistically significant, which makes them more credible.  A major implication of this result is that optimal classification results can be obtained without hyperparameter optimization, which  makes these models suitable for real-time use.

\begin{table}[H]
\centering
\resizebox{\columnwidth}{!}{
\begin{tabular}{llll}
\toprule
\multicolumn{4}{c}{\textbf{Execution Times (seconds)}} \\
\midrule
\textbf{Dataset} & \textbf{Split} & \textbf{Classifier} & \textbf{Time (s)} \\
\midrule
\multicolumn{4}{c}{\textbf{BODMAS 30/30 Split}} \\
\midrule
BODMAS & 30/30 & Random Forest        & 34.6131 \\
BODMAS & 30/30 & Logistic Regression  & 0.8388  \\
BODMAS & 30/30 & Naive Bayes          & 0.5964  \\
BODMAS & 30/30 & Decision Tree        & 3.9037  \\
BODMAS & 30/30 & LightGBM             & 1.6439  \\
\midrule
\multicolumn{4}{c}{\textbf{BODMAS 50/30 Split}} \\
\midrule
BODMAS & 50/30 & Random Forest        & 69.9131 \\
BODMAS & 50/30 & Logistic Regression  & 1.1087  \\
BODMAS & 50/30 & Naive Bayes          & 0.7727  \\
BODMAS & 50/30 & Decision Tree        & 7.2392  \\
BODMAS & 50/30 & LightGBM             & 2.7356  \\
\midrule
\multicolumn{4}{c}{\textbf{BODMAS 70/30 Split}} \\
\midrule
BODMAS & 70/30 & Random Forest        & 102.7147 \\
BODMAS & 70/30 & Logistic Regression  & 1.3402   \\
BODMAS & 70/30 & Naive Bayes          & 0.9787   \\
BODMAS & 70/30 & Decision Tree        & 10.7152  \\
BODMAS & 70/30 & LightGBM             & 5.5132   \\
\midrule
\multicolumn{4}{c}{\textbf{EMBER 30/30 Split}} \\
\midrule
EMBER  & 30/30 & Random Forest        & 173.0924 \\
EMBER  & 30/30 & Logistic Regression  & 2.0239   \\
EMBER  & 30/30 & Naive Bayes          & 1.4704   \\
EMBER  & 30/30 & Decision Tree        & 17.1457  \\
EMBER  & 30/30 & LightGBM             & 4.6313   \\
\midrule
\multicolumn{4}{c}{\textbf{EMBER 50/30 Split}} \\
\midrule
EMBER  & 50/30 & Random Forest        & 285.4887 \\
EMBER  & 50/30 & Logistic Regression  & 2.4814   \\
EMBER  & 50/30 & Naive Bayes          & 1.5534   \\
EMBER  & 50/30 & Decision Tree        & 30.7527  \\
EMBER  & 50/30 & LightGBM             & 6.3087   \\
\midrule
\multicolumn{4}{c}{\textbf{EMBER 70/30 Split}} \\
\midrule
EMBER  & 70/30 & Random Forest        & 420.7596 \\
EMBER  & 70/30 & Logistic Regression  & 2.9080   \\
EMBER  & 70/30 & Naive Bayes          & 1.8112   \\
EMBER  & 70/30 & Decision Tree        & 45.1684  \\
EMBER  & 70/30 & LightGBM             & 8.4585   \\
\bottomrule
\end{tabular}
}
\caption{Execution Times of Various Classifiers on the BODMAS and EMBER Dataset Splits}
\label{tab:execution-times}
\end{table}
The originality of this research is based on the fact  that this research shows how applying the concept of latent space can reduce the complexity of malware classification and at  the same time reduce the computational costs of the classification process. Unlike many other approaches that require much adjustment  of their hyperparameters, the method proposed in this paper produces stable classification results without optimization, which makes  it easy to use in a rapidly changing cyber security environment.

The main results of the work can be  summarized as follows. First, it has been found that ensemble methods can enhance the performance of malware classification  using latent space features and reduce the model size and computational time. Second, the use of statistical significance  testing is crucial in evaluating the classifier performance, to ensure the reliability and replicability of the results.  Third, it offers a basis for future work on the application of techniques for explaining the model’s  predictions and the application of deep learning models to enhance the quality of the  latent space for malware analysis.

Future work may include the continuation of the above research to establish how the  mentioned findings can be combined with interpretability techniques in order to gain a better understanding of the features influencing  the latent space. Also, the effectiveness of deep learning models on latent space features could be investigated with  the aim of enhancing malware detection. In summary, this paper proposes a malware classification method that is  efficient, reliable and easily expandable to other data sets, thus addressing the challenges of the current malware analysis  problems.

\FloatBarrier

\end{document}